\documentclass[aps,prd,twocolumn,nofootinbib,epsfig, showpacs]{revtex4}
\usepackage{amsmath,amssymb,latexsym}
\usepackage{graphicx}% Include figure files
\usepackage{bm}% bold math
\def\be{\begin{equation}}
\def\ee{\end{equation}}
\def\ed{\end{document}}

\begin{document}

\title{Confronting models on cosmic ray interactions with particle physics at LHC energies}
\author{Mar\'{\i}a Teresa Dova and Sergio Ferrari}
\affiliation{IFLP (CONICET-UNLP), Dpto. de F\'{\i}sica, Universidad Nacional La Plata, C.C.67 - 1900, La Plata, 
Argentina}

\begin{abstract}

Inelastic $pp$ collisions are dominated by soft (low momentum transfer) physics
where perturbative QCD cannot be fully applied. A deep understanding of both
soft and semi-hard processes is crucial for predictions of minimum bias and
underlying events of the now coming on line $pp$ Large Hadron Collider (LHC).
Moreover, the interaction of cosmic ray particles entering in the atmosphere is
extremely sensitive to these soft processes and consequently cannot be
formulated from first principles. Because of this, air shower analyses strongly
rely on hadronic interaction models, which extrapolate collider data several
orders of magnitude. A comparative study of Monte Carlo simulations of $pp$
collisions (at the LHC center-of-mass energy $\simeq 14~{\rm TeV}$)
using the most popular hadronic interaction models for ultrahigh energy cosmic
ray (SIBYLL and QGSJET) and for collider physics (the PYTHIA multiparton model)
is presented. The most relevant distributions are studied including those
observables from diffractive events with the aim of discriminating between the
different models.

\end{abstract}

\maketitle

\section{Introduction}

Particle colliders and ultra high energy cosmic ray (UHECR) experiments are at present the best scenarios to test the deep structure of matter and the interactions of its fundamental constituents at the frontier of energy. To this end, detailed modeling of the underlying physics, based on simulation programs also known as event generators, are required. These models are very important tools to define experimental and analysis strategies, to test new theoretical ideas and to design new experiments. 

From the perspective of particle physics, UHECR interactions are orders of magnitude beyond what can be achieved in current (and future) terrestrial collider experiments and may open a window to energy and kinematic regions previously unexplored in the study of fundamental interactions.  The Pierre Auger Collaboration, exploding the potential of the hybrid design of the observatory (ground based and fluorescence detectors) has developed a method to obtain the energy spectrum which does not rely on detailed numerical simulations extrapolated from experimental knowledge of man-made accelerators to the highest cosmic ray energies \cite{sommers}. The hybrid detection provides a way to inter-calibrate the subsystems and to control systematic uncertainties \cite{nim}. This new approach to derived the cosmic ray spectrum will allow to constrain, although in an indirect manner, interaction models at energies and phase space regions that complement those of colliders experiments. For the case of primary composition determination, however, UHECR data are interpreted in all cases using Monte Carlo simulations~\cite{Anchordoqui:1998nq,agasa, hires, knapp-us}.

The Large Hadron Collider (LHC) \cite{LHC}, currently under construction at CERN, will provide $ pp$ collisions at the unprecedented centre-of-mass energy of $\sqrt{s}$ = 14 TeV and luminosity of ${\cal{L}}=10^{34}$ $\rm{cm}^{-2} \rm{s}^{-1}$. It will also provide heavy ion collisions at a centre-of-mass energy of about 1000 TeV. A good description of both hard and soft processes in event generators at LHC energies are very important to understand the {\it underlying event} associated with high transverse momentum ($p_{T}$) jets or leptons as well as {\it minimum bias events}. These events will be responsible for most of the radiation background expected at LHC experiments and hence it is essential to study detector damages, triggering systems, detector counting rates, etc.  {\it Minimum bias events} are worthy os scientific study as they provide a good insight into the internal structure of protons. In addition, dedicated runs of the LHC with lower luminosity (${\cal{L}} = 10^{28}$$ \rm{cm}^{-2} \rm{s}^{-1}$) and specially tuned beam optics are planned to study diffractive events. Both ATLAS\cite{ATLAS} and CMS\cite{CMS} experiments are planning to implement additional detectors to cover the forward diffractive regions with tracking and/or calorimetry \cite{D13,TOT,ATLASDIF}. Recently, a new experiment to study very forward particle emission in the LHC collider, LHCf, has ben approved. The experimental results of LHCf will be able to provide the production spectum of secondary particles in the very forward region, allowing to constrain the Monte Carlo codes \cite{LHCf}. Moreover, novel aspects of diffraction studies are included in the physics case for forward proton tagging at 420 m at the LHC \cite{FP420}.  A study of diffraction must use detectors with excellent forward acceptance to allow for a comparison with cosmic ray data.  A good review of diffractive physics can be found in ref.\cite{PED}, while references \cite{D1,D3,D14,D17} focus on future studies at LHC.

In this paper we present a comparative study between the two most frequently applied models  for simulation of cosmic rays extensive air showers SIBYLL \cite{SIB} and QGSJET \cite{QGS} with a multi-purpose Monte Carlo like PYTHIA \cite{PYT} tunned for use in LHC experiments.  The paper is organized as follows: In section II the main features of the models used in the Monte Carlo generators for $pp$ collisions with emphasis in the most distinctive differences among them are presented. The method and the results from the Monte Carlo studies  at LHC center of mass energy with the aim of discriminating between the different models are described in section III with two parts. In the first part, predictions for the most relevant distributions in studies of both collider and UHECR data are discussed. In the second part a comparative analysis of the signatures from diffractive events is presented. Final comments and conclusions are given in section IV.

\section{Models for hadronic collisions}

Although electromagnetic and weak interactions are well understood, this is not the case for hadron production in collisions of nucleons, pions and kaons with light nuclei, where the lack of experimental data posses limitations in many UHECR and accelerator applications \cite{engel0504358,knapp-us}. This is mainly because precise perturbative quantum chromo-dynamics (pQCD) calculations are only possible for processes with large momentum transfer, also known as ``hard'' interactions, that constitute only a minute fraction of the overall reaction rate. In hadron-hadron interactions it is customary to distinguish between elastic and inelastic processes, and this later into diffractive (including single and double diffraction) and non-diffractive ones (usually called {\it minimum bias events}). Precisely, low-$p_{T}$ (``soft'') processes, where pQCD cannot be fully applied and phenomenology models are used,  play a dominant role in the non-diffractive component. 

Current models of high energy hadron collisions  typically rely on the pQCD formalism in the description of high-$p_{T}$ scattering, while treating the low-$p_{T}$ ones in a simplified phenomenological approach. At the LHC, the description of the new physics processes to be studied are mostly controlled by pQCD. Some efforts have been done to investigate the models used by the most popular collider event generators, like PYTHIA, to describe ``soft'' interactions in hadron-hadron collisions with the aim of predicting {\it minimum bias } and   {\it underlying event} levels of particle production at the LHC \cite{MB1,MB2,MB3, MB4, MB5, MB6}. These studies resulted in a tuning of PYTHIA based on comparisons of experimental data that is used in this paper. 

The pQCD inclusive cross section of production of parton jets pairs with transverse momenta larger than some cutoff $Q_{\rm min}^2$ is given by,
\begin{widetext}
\begin{equation}
\sigma_{\rm QCD}(s,p_{{_T}}^{\rm cutoff}) = \sum_{i,j} \int
\frac{dx_1}{x_1}\int \frac{dx_2}{x_2} 
  \int_{Q_{\rm min}^2}^{\hat{s}/2}d|\hat t| 
\frac{d\hat{\sigma}_{ij}}{d|\hat t|}
x_1 f_i(x_1, |\hat t|)  x_2 f_j(x_2, |\hat t|) \,\,,
\label{sigmaminijet}
\end{equation}
\end{widetext}
where $x_1$ and $x_2$ are the fractions of the momenta of the parent hadrons
carried by the partons which collide,
$d\hat{\sigma}_{ij}/d|\hat t|$ is the cross section for scattering of
partons of types $i$ and $j$ according to elementary QCD diagrams,
$f_i$ and $f_j$ are parton distribution functions (pdf's),
$\hat{s} = x_1\,x_2 s$
and $-\hat{t} = \hat{s}\, (1 - \cos \vartheta^*)/2 =  Q^2$
are the Mandelstam variables for this parton-parton process,
and the sum is over all parton species.

In the UHECR field, the required information to model the interaction of the primary particle entering in the atmosphere appears to be extremely sensitive to the underlying ``soft'' non perturbative hadronic process \cite{OstapchenkoJPG:2003}. In this direction, there are three event generators, SIBYLL\cite{SIB}, QGSJET\cite{QGS}  and DPMJET\cite{DPMHET} which are tailored specifically for simulation of hadronic interactions up to the highest cosmic ray energies. 

The most frequently used high energy hadronic models in the study of UHECRs are QGSJET and SIBYLL. In these codes, the low $p_T$
interactions are modeled by the exchange of Pomerons.
Regge singularities are used to determine the momentum
distribution functions of the various sets of constituents,
valence and sea quarks. Both QGSJET and SIBYLL share the eikonal model
and then assume the unitarized cross sections assuming a real eikonal
function sum  of a soft and hard contributions:
\be
\sigma_{inel} = \int d^2 \vec{b}
 \left( 1 - \exp \{-2\chi_s(s,\vec{b}) - 2\chi_h(s,\vec{b})  \} \right)
\ee
At high energies the hard eikonal is dominating:
\be
\chi_h = \frac{1}{2} \sigma_{QCD}(s, p_T^{\mbox{cutoff}}) A(s,\vec{b}),
\ee
where the normalized profile function $A(s,\vec{b})$ describes the
distribution of partons in the plane transverse of the collision axis.
QGSJET and SIBYLL take different assumptions on the profile function
which determines the inelastic cross section and its energy dependence.\\
QGSJET assumes a gaussian profile distribution and its theory is formulated entirely in terms
of Pomeron exchanges. The basic idea is to replace the
soft Pomeron by a so-called semihard Pomeron, defined to be an ordinary soft Pomeron with the middle
piece replaced by a QCD parton ladder. Thus, minijets
will emerge as a part of the   semihard  Pomeron, which is
itself the controlling mechanism for the whole interaction.

%After performing the energy sharing among the soft and semihard Pomerons, and also the sharing among the soft and hard pieces of the last one; the number of charged particles in the partonic cascade is easily obtained generalizing the method of multiple production of hadrons as discussed in the QGS model (soft Pomeron showers).

In SIBYLL the profile function is based on the Fourier transform
of the electromagnetic form factor and it is an energy-independent
exponential.  The underlying idea behind SIBYLL
 is that the increase in the cross section is driven by the
 production of minijets. The probability distribution
 for obtaining N jet pairs (with $p_T^{\rm{jet}} >
 p_T^{\rm{min}}$, being $p_T^{\rm{min}}$ a sharp threshold on the
 transverse momentum below which hard interactions
 are neglected) in a collision at
energy $\sqrt{s}$ is computed regarding elastic $pp$
or $p\overline{p}$ scattering as a diffractive shadow
scattering associated with inelastic processes. The
algorithms are tuned to reproduce
the central and fragmentation regions data up to
$p\overline{p}$ collider energies, and with no further adjustments they
are extrapolated several orders of magnitude.

A general update of QGSJET has been recently presented, where the key improvement is connected to an account for non-linear interaction effects in individual hadronic collisions  \cite{QGSII}. Additionally, a more reliable low mass diffraction  treatment has been used and all model parameters have been re-calibrated using a wider set of accelerator data. This results in a new model, QGSJET-II. Non-linear screening corrections appear to be correlated with corresponding parton densities and become larger at higher energies, smaller impact parameters, resulting in the saturation of pdf's at the scale $Q_{\rm min}^2$ and in a considerable reduction of ``soft'' particle production. 

In the case of PYTHIA, perturbative QCD is used 
extending it for the case of low-pt. pQCD  is
divergent for  $p_T \to 0$, PYTHIA avoids the divergence
using two different scenarios.The ``simple scenario'' consists
in fixing a minimum value of $p_T^{\rm{min}}$ below which the cross section
is defined as null and can be interpreted as the inverse of some color
screening length in the hadron. This is equivalent to set
 a maximum impact parameter $b_{\rm{max}}$ above which there
is no more interaction. In the so called ``complex scenario''
a regulating parameter $p_{T0}$ is introduced 
below which the cross sections are dumped. Different models of matter distribution in the hadron are considered: uniform, simple gaussian and
double gaussian.
 
The transition process from asymptotically free partons to colour-neutral hadrons is describes in all codes by string fragmentation models \cite{ref102}.

In summary, there are differences between the models for hadronic collisions in the existing event generators that will emerge in the Monte Carlo study presented in the rest of this paper.

\section{Description of the method and Monte Carlo studies}

For analyzing the difference between models, we have generated
samples of $10^4$ $pp$ collisions at the LHC
center-of-mass energy for each event generator model: QGSJET-II \cite{QGSII}, QGSJET-01 \cite{QGS01}, SIBYLL 2.1 \cite{SIB21} and PYTHIA 6.205 \cite{PYT}. All calculations contain a mixture of diffractive and non-diffractive events according to the model used. All secondary particles were registered without any energy cut.

In the case of PYTHIA, it was chosen the ``complex scenario''
with a double gaussian distribution of matter inside the hadron. The default values of some parameters were modified according to the results of references
 \cite{MB1,MB2,MB3, MB4, MB5, MB6} where the optimal values where obtained from a tunned PYTHIA using events from different experiments. In the following table both default and tunned values of PYTHIA as used in this paper are presented:
\begin{table}[h!]
\caption{Values of relevant PYTHIA 6.2 parameters.}
\begin{tabular}{|l|c|c|}
  \hline
Variable      &  Default &  Tunned \\
\hline
MSTP(81)  &  1  &  1  \\
MSTP(82)  &  1  &  4  \\
PARP(82)  & 2.1 & 1.8 \\
PARP(83)  & 0.5 & 0.5 \\
PARP(84)  & 0.2 & 0.5 \\
MSTP(2)   &  1  &  1   \\
%MSTP(33)  &  0  &  0   \\
\hline 
\end{tabular}
\end{table}
where MSTP(81) refers to the master switch for multiple interactions,
MSTP(82) = 4 selects an hadronic matter
overlap consistent with a given  double Gaussian matter distribution
 and a continuous turn-off of the cross
section at $p_{T0}$ =PARP(82). This double Gaussian matter
distribution is regulated by the following parameters: a core
 PARP(84) of the main radius containing a fraction
PARP(83) of the total hadronic matter. The value of the
parameter MSTP(2) gives the kind of  calculation of $\alpha_S$
at hard interaction and if the value is 1 then it is first-order
running of $\alpha_S$ (Here $\alpha_S$ is the strong coupling constant) 

\begin{figure}
\centering 
\includegraphics[width=8.cm, height=8.cm]{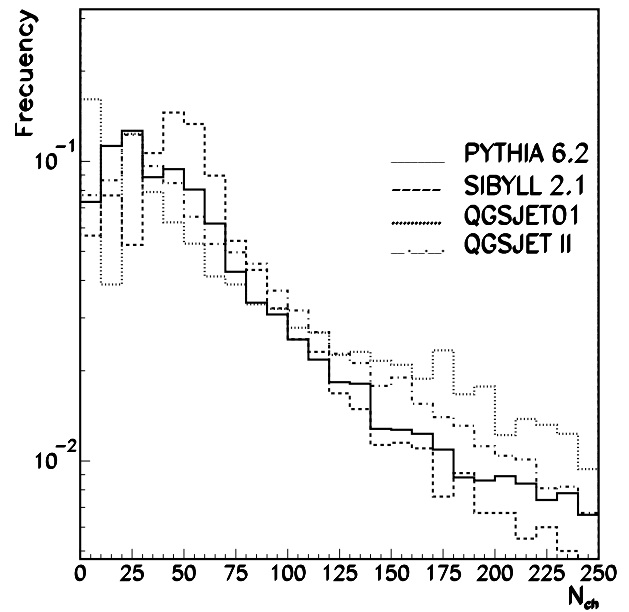}
\caption{Distribution of the number of charged secondaries produced
in $pp$ collision at LHC energy. The solid line
corresponds to PYTHIA 6.2, dotted to QGSJET-01, the dashed one to SIBYLL2.1 and dashed-dotted to QGSJET-II}\label{nsec} 
\end{figure}

\subsection{General features}

To help understanding the differences of the models currently used in the study of UHECRs, when extrapolated from collider data to higher energies and to small angle processes, it is important to compare those variables driving the development of air showers, i.e, the multiplicity of produced secondary  particles and the inelasticity (the relative energy loss of leading secondaries).
  
Multiplicity distributions of charged secondaries
($N_{ch}$) produced in each collision are shown in
Fig.\ref{nsec}. Table 2 shows the average number of secondaries for each model. Besides the mean values, the standard deviations over the $10^4$ interactions are given. The errors of the mean values ($\sigma /\sqrt(10^4)$) are much smaller. SIBYLL produces 60\% to 65\% less nucleons than the other models. This has been noticed in previous analysis at higher energies \cite{knapp-us,Anchordoqui:1998nq} and it is also seen at LHC energy. QGSJET-01 produces more baryons and charged pions than the other models, but this situation has changed in QGSJET-II, due to the non-linear screening corrections, which lead to a reduction of the interaction eikonal and hence of the number of elementary particle production processes \cite{QGSII}. This results in a reduction of particle production in QGSJET-II compared with QGSJET-01, bringing it closer to SIBYLL and in good agreement with the PYTHIA predictions. The mean charged particle multiplicity, which already shows differences between models at this energy, readily increases with rising energy as QGSJET predicts  a power law-like increase of the number of secondaries, while SIBYLL  multiplicity exhibits a logarithmic growth and PYTHIA follows a $\ln ^{2}(s)$. It is worth mentioning here the results reported by the CDF collaboration favouring an energy dependence stronger than  $\ln (s)$ \cite{Abe}.

\begin{table}
\caption{Average multiplicity and inelasticity per proton-proton collision.}
\begin{tabular}{|c|c|c|c|c|}
\hline
Variable      &  PY 6.2    &  QGS 01   & QGS II     & SIB 2.1 \\
\hline
$p$            & 3.8  (3.1) & 3.5 (2.7) & 3.9 (3.0)  & 2.6 (1.7)\\
$\overline{p}$ & 2.5  (3.1) & 2.3 (2.2) & 2.6 (2.9)  & 1.2 (1.6)\\
$n$            & 5.6  (5.9) & 5.3 (5.2) & 5.7 (5.7)  & 3.2 (3.1)\\
$\pi^\pm$      & 66.5 (72.1) & 70.2 (68.3) & 66.9 (64.5) & 64.7 (60.8)\\
$\pi^0$        & 37.0 (40.4) & 35.9 (34.9) & 34.7 (33.7) & 38.9 (37.2) \\
$K^\pm$        & 7.5 (8.9)  & 9.9 (9.9) &  6.8 (6.9) & 7.6 (8.1) \\
$K_L$          & 3.6 (4.5)  & 4.9 (5.1) &  4.4 (3.7) & 3.7 (4.2) \\
\hline
$N_{charged}$  & 80.3 & 85.9  & 80.3& 76.1 \\
\hline 
$N_{total}$    & 126.5 & 139.3 & 136.1 & 125.7 \\
\hline
$<k_L>$  &  0.41 & 0.50 & 0.43 & 0.43\\
\hline
\end{tabular}
\end{table}

\begin{figure}
\centering 
\includegraphics[width=8.cm, height=8.cm]{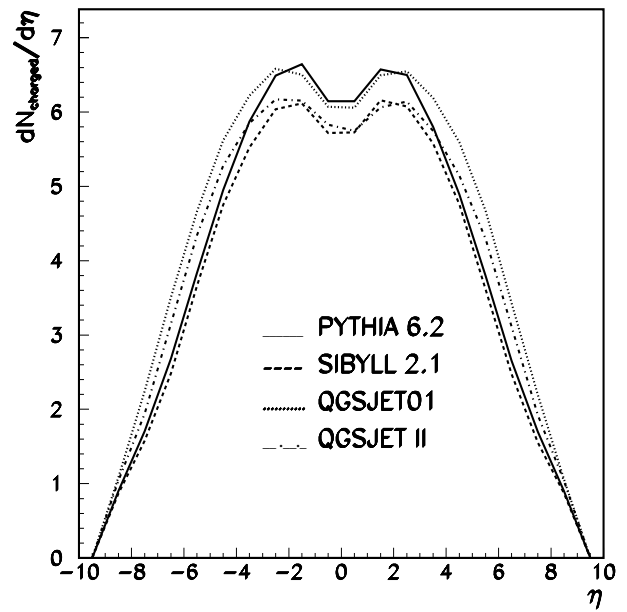}
\caption{Pseudorapidity distribution for charged particles.}\label{pcharm} 
\end{figure}

The distribution in pseudorapidity, $\eta = - \ln \tan(\theta/2)$, of charged particles for $pp$ collisions at 14 TeV is presented
in Fig. \ref{pcharm}.  It shows up clearly that QGSJET-01 produces
more secondaries than the other models in all directions but in the central region of $|\eta | $, where it is superseded by PYTHIA. This might be explained by the pure QCD treatment and the possibility of multiple interactions set in PYTHIA. It has been noticed \cite{MB2} that this tunned version of PYTHIA provides the best description of experimental data from UA5 and CDF in the central rapidity region. Above $|\eta | > 3 $ PYTHIA and SIBYLL give similar predictions, both having smaller values than the QGSJET models.

\begin{figure}
\centering 
\includegraphics[width=8.cm, height=8.cm]{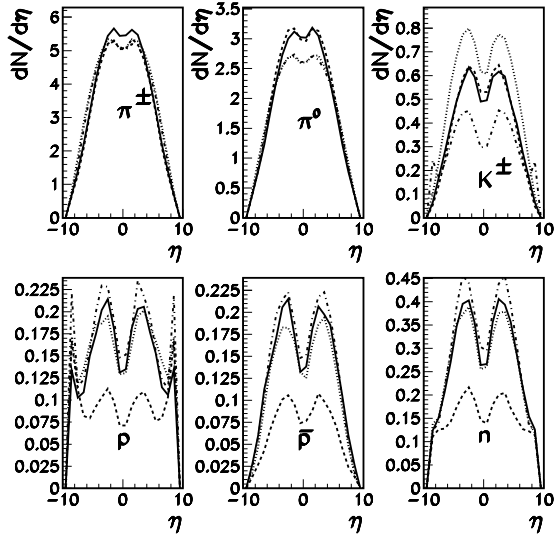}
\caption{Pseudorapidity distribution for different types of particles. Left-top panel corresponds to charged pions, right-top to neutral pions, left-bottom to protons and right-bottom panel corresponds to neutrons. The solid line
corresponds to PYTHIA 6.2 events, dotted to QGSJET-01, the dashed one to SIBYLL2.1 events, dashed-dotted to QGSJET-II.}\label{ppartm} 
\end{figure}

  To get additional information, the pseudorapidity distributions for charged and neutral pions, kaons, protons, antiprotons and neutrons are shown in Fig. \ref{ppartm}.  There are discrepancies between the models in all cases being the largest ones for nucleons and antiprotons, where although models predict similar shapes, SIBYLL exhibits a clear deficit at all $ \eta $. For the case of kaons there is a factor of two in the predictions from the two versions of QGSJET. The peaks in the very forward and backward parts of the pseudorapidity distribution for protons correspond to diffractive events, in which one of the smashing protons keeps traveling approximately in the same direction after the collision. The shoulders in the high pseudorapidy region for neutrons have a different origin. If a neutron comes out as the fast particle, charge is being exchanged. The process, accordingly, cannot be attributed to the exchange of zero quantum numbers (i.e to diffraction) but, for instance, to pion exchange. It is worth noting that the differences in the production of neutral pions influence the shower development of the secondary particles produced by the interaction of a primary cosmic ray particle in the atmosphere, which is driven by the electromagnetic component generated from the $\pi^{0}$s.  

%\begin{figure}
%\centering 
%\includegraphics[width=15.cm, height=10.cm]{fig3b.eps}
%\caption{Pseudorapidity distribution for different types of particles. Left-top panel are charged pions, right-top are neutral pions, left-bottom are protons and right-bottom panel are neutrons.The solid line
%corresponds to PYTHIA 6.2 events, dotted to QGSJET-01, the dashed one to SIBYLL2.1 events, dashed-dotted to QGSJET-II}\label{ppartm} 
%\end{figure}

\begin{figure}
\centering 
\includegraphics[width=8.cm, height=8.cm]{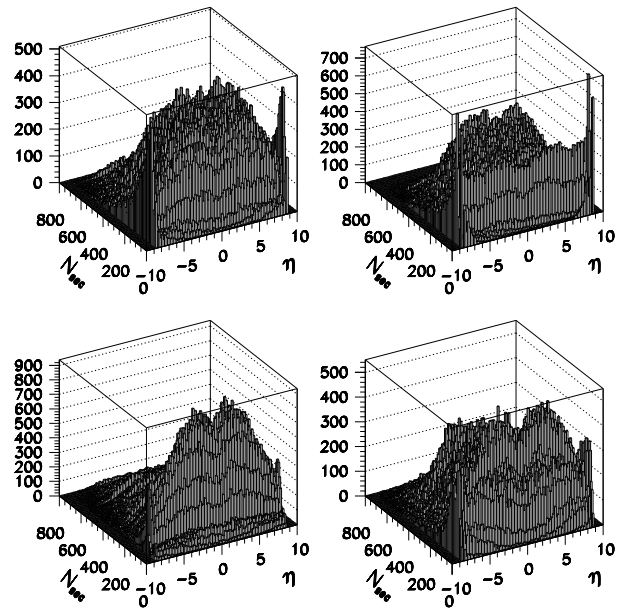}
\caption{Number of secondaries vs pseudorapidity 2D-distribution. Top-left panelcorresponds to QGSJET-II, top-right panel corresponds to QGSJET-01,
bottom-left panel to SIBYLL 2.1 and bottom-right panel
to PYTHIA 6.2. }\label{2detasec}
\end{figure}

Two-dimension distributions of number of secondaries, $N_{sec}$, vs pseudorapidity, $\eta $, are presented in Fig \ref{2detasec}.  The plots show in detail  the large differences between QGSJET, upper panels, and SIBYLL, bottom-left panel,  in the whole region of $\eta $. The two diffractive peaks in the region of low multiplicity and high
pseudorapidity are well separated from the broad distribution of non-diffractive events in both versions of QGSJET and to less extent in PYTHIA. This feature is due to the fact that QGSJET models have none or few non-diffractive events with small multiplicities, while in PYTHIA  the distributions overlap and diffractive events tend to have higher multiplicities as well. For the  SIBYLL model the distribution for low number of secondaries is rather flat in all directions.  

Table III shows the percentage frequency of the leading particle produced in the collision. In more than 50\% of the collisions protons emerge as these leading particles. In SIBYLL an almost 65 \% the most energetic particle is a proton, while the other models give between 43\% to 62\%. SIBYLL and PYTHIA generate mesons as the most energetic particle in $\approx 20\% $ of the cases, however QGSJET-01 and QGSJETII have larger and smaller productions of mesons as leading particles respectively with a difference of roughly a factor of 3. All models assume that the leading particle distributions scale with energy, being tunned to low energy. Certainly, measurements of hadron production in the very forward region are needed to study the leading baryon distributions, mainly because there are some theoretical models predicting that the leading particle distributions will change drastically at very high energies \cite{engel0504358,frankfurt}

 \begin{table}
\caption{Most energetic secondary particle probabilities.}
\begin{tabular}{|c|c|c|c|c|}
\hline
            & Pythia 6.2 &  Qgsjet 01  & Qgsjet II & Sibyll 2.1 \\
\hline
proton      &    55.29\%  &  43.27\%    &   62.08\% &   64.62\%  \\
neutron     &    27.34\%  &  18.31\%    &   19.68\% &   16.51\%  \\
\hline
$\Sigma$ nucleons &  82.63\%  &  61.58\%    &   78.76\% &   81.13\%  \\
\hline
$\pi^{\pm}$ &    10.28\%  &  20.47\%    &    6.77\% &   10.72\%  \\
$\pi^0$     &    4.89\%   &   9.74\%    &    3.02\% &   5.85\%  \\
$K^\pm$     &    1.57\%   &   2.40\%    &    0.73\% &   1.00\%  \\
$K_L$       &    0.63\%   &   0.91\%    &    0.44\% &   0.57\%  \\
\hline
\end{tabular}
\end{table}

\subsection{Signatures of diffractive events}

As mentioned above, in hadron-hadron interactions the inelastic processes are usually divided into diffractive and non-diffractive. In this section, a study of the predicted signatures from different models for diffractive hadronic interactions is presented.

\begin{figure}
\centering 
\includegraphics[width=8.cm, height=8.cm]{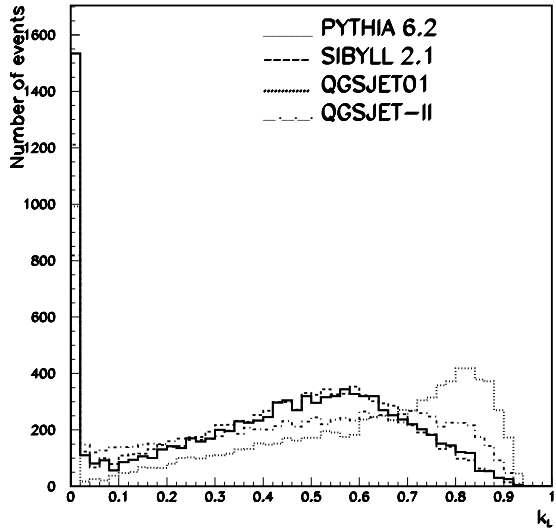}
\caption{Inelasticity ($k_L = 1 - \frac{E_{\rm{lead}}}{E_{P}}$) distriution. The solid line
corresponds to PYTHIA 6.2, dotted line to QGSJET-01, dashed line to SIBYLL2.1 and  dashed-dotted one to QGSJET-II .}\label{inel} 
\end{figure}

A good parameter for tangling diffractive events from $pp$ collisions is the inelasticity defined as:
\be
k_L = 1 - \frac{E_{\rm{lead}}}{E_{P}}
\ee
 where $E_{P}$ is the energy of the incident particle in the lab frame, and
$E_{\rm{lead}}$ is the energy of the secondary with largest
energy (the so-called leading particle). A signature that can be used to distinguish diffractive from  non-diffractive events is their low value in both 
inelasticity and number of secondaries \cite{IDICRAS}. In Table II the value of
 the average
 inelasticity for each model is shown, while  the corresponding  inelasticity distributions are displayed in Fig. \ref{inel}. A narrow peak at low $k_L$  is evident from this plot, which corresponds to elastic and single diffractive processes. For non-diffractive events, the available energy is shared among many secondaries leading to a rather uniform distribution in the whole range of $k_L$.

A few differences between models can
be seen in the inelasticity distribution: PYTHIA has the highest
diffractive peak. QGSJET-01 also presents a small peak
at large $k_L $ due to the large number of secondaries produced in inelastic collisions. QGSJET-II has no longer that feature. SIBYLL and PYTHIA distributions are in general good agreement.

\begin{figure}
\centering 
\includegraphics[width=8.cm, height=8.cm]{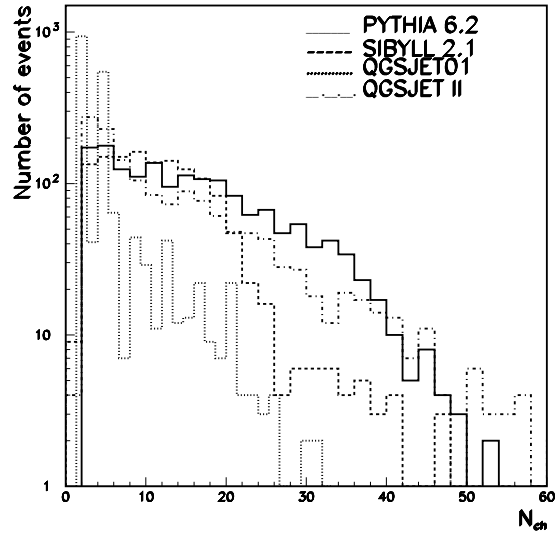}
\caption{Distribution of the number of secondaries for events with
low inelasticity ($k_L \le 0.04$).}\label{nsklow} 
\end{figure}

\begin{figure}
\centering 
\includegraphics[width=8.cm, height=8.cm]{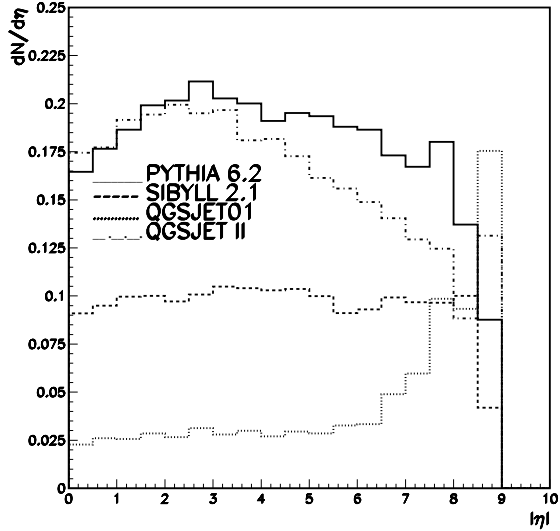}
\caption{Pseudorapidity distribution for the selected ``diffractive''events. }\label{diff} 
\end{figure}

A close inspection of the multiplicity for single diffractive, double diffractive and non-diffractive events using PYTHIA indicates that cutting at multiplicity below 40 the diffractive events largely dominates the sample. However, a cut in low inelasticity ($k_L < 0.04$) leave an even more pure diffractive sample.  We will then label as ``diffractive'' events
the ones with inelasticity $k_L < 0.04$. In Fig. \ref{nsklow} the distribution of charged particles, $N_{ch}$, for $k_L < 0.04$ is presented. Both QGSJET-01 and
SIBYLL distributions barely goes further than 40 secondaries, while QGSJET-II and PYTHIA extend up to 60 secondaries.  Figure  \ref{diff} shows particle densities distributed in pseudorapidity space for ``diffractive'' events. There are large divergencies between PYTHIA and SIBYLL in the predictions of the particle multiplicity in the whole region of $\eta$. PYTHIA predicts a density roughly 80 \% greater than SIBYLL. QGSJET-01 shows a rather flat distribution at intermediate values between PYTHIA and SIBYLL, while the new version QGSJET-II is in good agreement with PYTHIA at $|\eta|<4$ and present smaller values at  $|\eta|>4$. It is worth mentioning here that the cuts $k_L < 0.04$ and  $N_{sec}<40$ in the PYTHIA sample allow selection of a pure sample of single diffractive events.

\begin{figure}
\centering 
\includegraphics[width=8.cm, height=8.cm]{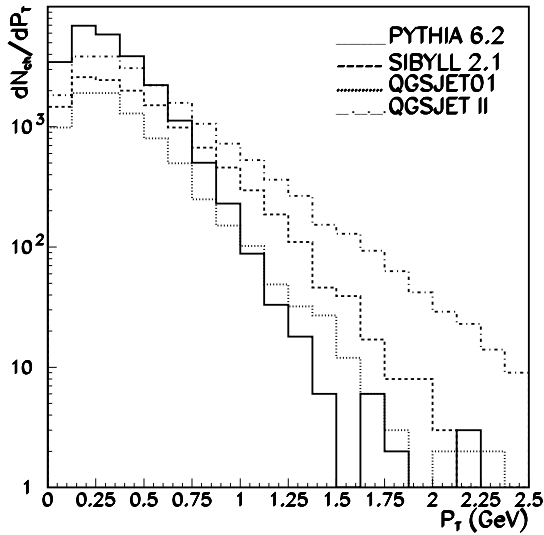}
\caption{$p_T$ distribution for the selected ``diffractive''events. }\label{diffpt} 
\end{figure}

The transverse momentum, $p_T$, detection capabilities at ATLAS and CMS  with good resolution will be limited to particles with $p_T > 0.5 $ GeV \cite{ATLAS,CMS} where for the labeled ``diffractive'' events QGSJET-II and PYTHIA show the largest difference. This is evident in Fig.  \ref{diffpt} where the $p_T$ distribution of eventsin the central region ( $|\eta|<5$) is shown. At low momenta $dN_{\rm{ch}}/dp_T$ is greater for PYTHIA, but as $p_T$ increases densities for the other models become greater with a difference of an order of magnitud at  $p_T > 1.5 $ for the case of QGSJET-II, as PYTHIA does not create diffractive events with high transverse momemtum.


\begin{figure}
\centering 
\includegraphics[width=8.cm, height=8.cm]{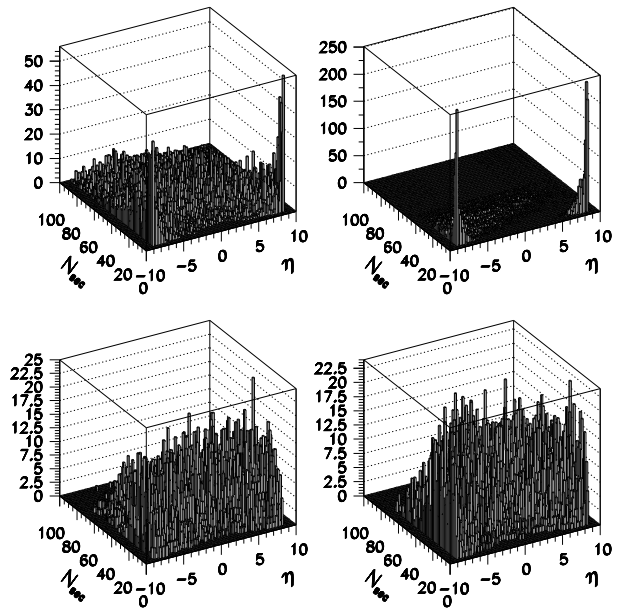}
\caption{Multiplicity vs pseudorapidity 2D-distribution for
low inelasticity ($k_L \le 0.04$) events. Top-left panel corresponds to QGSJET-II,
top-right panel corresponds to QGSJET-01, bottom-left panel to SIBYLL 2.1 and bottom-right panel to PYTHIA 6.2. }\label{2dlow}
\end{figure}

Figure  \ref{2dlow} shows 2-D distributions of the number of secondaries in the labeled ``diffractive'' events vs. pseudorapidity. Again, it is evident that the distributions generated by SIBYLL are fundamentally different to the ones from QGSJETs and PYTHIA. This is certainly due to the phenomenological description of diffractive events in SIBYLL \cite{SIB}. QGSJET-01 and QGSJET - II  predict a large fraction of events with low number of secondaries in the region of high pseudorapidity. Clearly, the better treatment of diffraction for the case of QGSJET-II with its parameters tunned using accelerator data, results in diffractive peaks much lower than in QGSJET-01, but still larger compared with PYTHIA data.

\section{Conclusions}

The two most widely used high energy hadronic interaction models for the study of cosmic rays, SIBYLL and QGSJET, and the most popular simulation program for collider experiments, PYTHIA, have been compared at LHC center of mass energy. These codes using different theoretical models involving many variables have different predictions for the most important observables, such as particle multiplicities, distribution of particles in pseudorapidity space and transverse momentum distribution, allowing to investigate, and improve, the theoretical predictions of hadron-hadron interactions at this energy.

The discrepancies observed in the  pseudorapidity distributions for different types of particle, in particular in the acceptance region of LHC experiments would allow to perform a combined analysis to constrain the models, using the large {\it minimum bias events} statistics that will be collected in the very first LHC operation. The predictions for the 2-D distributions of particle multiplicity vs pseudorapidity will also help deciding on the best theoretical model.

 Measurements of hadron production with the forward detectors attached to LHC experiments are very important for a better understanding of the simulations that model soft hadronic interactions at high energies. The analysis of the percentage frequency of the leading particle produced in the collision indicates differences up to a factor of 2 for meson production. 

A selection of events from $pp$ collisions at $\sqrt s=14 \rm{TeV}$ with small inelasticity ($k_L < 0.04$) and low number of secondaries allows to pick diffractive events for a comparative study of the various models. The analysis of particle densities in pseudorapidity space indicates a good agreement of predictions using PYTHIA and QGSJET-II and a clear deficit of particle densities using SIBYLL and to less extent QGSJET-01.  The $p_T$ distributions in the central pseudorapidity region clearly indicate that SYBYLL and QGSJET models create up to an order of magnitude more particles with large $p_T$ than PYTHIA. A comparison with experimental data will provide strong constrains in modeling diffractive physics.
     
 The lack of suitable accelerator data is the dominant source of systematic uncertainties in the  analysis of the extensive air shower data. At the same time, analysis of {\it minimum bias events} at LHC are very important for understanding the {\it underlying event} and commissioning studies for LHC detectors. Certainly, the discrepancies in the models discussed in this paper will be naturally reduce with the large statistics of interesting data at a completely new energy frontier for terrestrial colliders, as LHC, and cosmic ray experiments, as the Pierre Auger Observatory.

\section*{Acknowledgments}
We would like to thank S. Ostapchenko for illuminating discussions to
understand the new features on QGSJET-II and providing an early
version of the program code. Special thanks to A. De Roeck for
bringing up our interest in the subject and to L. Anchordoqui for
valuable comments on the manuscript.

\end{document}